\newcommand{\version}{December 18, 2009}
\documentclass[a4paper,11pt]{article}
\usepackage{amsmath}
\usepackage{amsfonts}
\usepackage[bindingoffset=0.5cm,textheight=22.5cm,hdivide={2.9cm,*,2.9cm}, vdivide={*,22cm,*}]{geometry}
\usepackage[dvips]{graphicx}
\usepackage[dvips,bookmarksnumbered=true,breaklinks=true]{hyperref}
\usepackage[square,comma,sort&compress]{natbib}
\newcommand{\figref}[1]{Fig.~\ref{#1}}			% for figures
\newcommand{\inv}[1]{\frac{1}{#1}}
\newcommand{\tinv}[1]{\tfrac{1}{#1}}
\newcommand{\starco}[2]{\left[#1\stackrel{\star}{,}#2\right]}
\newcommand{\staraco}[2]{\left\{#1\stackrel{\star}{,}#2\right\}}

\newcommand{\var}[2]{\frac{\d #1}{\d #2}}
\newcommand{\vvar}[3]{\frac{\d^2 #1}{\d #2\d #3}}

\newcommand{\intx}{\int d^4x}
\newcommand{\intk}{\int d^4k}
\newcommand{\Gam}{\Gamma^{(0)}}
\newcommand{\Act}{S}
\newcommand{\pa}{\partial}
\newcommand{\ri}{{\rm i}}
\newcommand{\re}{{\rm e}}
\newcommand{\ig}{{\rm i}g}
\renewcommand{\a}{\alpha}
\renewcommand{\b}{\beta}

\newcommand{\e}{\epsilon}
\newcommand{\s}{\sigma}
\renewcommand{\d}{\delta}

\renewcommand{\th}{\theta}
\renewcommand{\e}{\epsilon}
\newcommand{\vare}{\varepsilon}

\newcommand{\W}{\Omega}
\renewcommand{\k}{\tilde{k}}
\newcommand{\p}{\tilde{p}}
\newcommand{\q}{\tilde{q}}
\newcommand{\wtild}{\widetilde}
\newcommand{\bc}{\bar{c}}
\newcommand{\tc}{\widetilde{c}}

\newcommand{\tx}{\tilde{x}}

\newcommand{\uim}{UV/IR mixing}
\newcommand{\nc}{non-commutative}
\newcommand{\dif}[2]{\frac{\delta#1}{\delta#2}}
\newcommand{\x}{\tilde{x}}

\newcommand{\Ga}{\Gamma_{\text{tot}}}
\title{\begin{flushright}
       \small{TUW-09-21}\\\small{UWThPh-2009-15}
       \end{flushright}
\vspace{2em}
Loop Calculations for the Non-Commutative \texorpdfstring{$U_\star(1)$}{U(1)} Gauge Field Model with Oscillator Term}

\author{Daniel N. Blaschke\footnotemark[1]~\footnotemark[2]~, Harald Grosse\footnotemark[2]~, Erwin Kronberger\footnotemark[1]~, \\Manfred Schweda\footnotemark[1]~ and Michael Wohlgenannt\footnotemark[1]}

\date{\version}

\graphicspath{{figures/}}

\begin{document}
\maketitle
\thispagestyle{empty}
\begin{center}
\renewcommand{\thefootnote}{\fnsymbol{footnote}}
\vspace{-0.3cm}\footnotemark[1]Institute for Theoretical Physics, Vienna University of Technology\\Wiedner Hauptstrasse 8-10, A-1040 Vienna (Austria)\\[0.3cm]
\footnotemark[2]Faculty of Physics, University of Vienna\\Boltzmanngasse 5, A-1090 Vienna (Austria)\\[0.5cm]
\ttfamily{E-mail: blaschke@hep.itp.tuwien.ac.at, harald.grosse@univie.ac.at, kronberger@hep.itp.tuwien.ac.at, mschweda@tph.tuwien.ac.at, miw@hep.itp.tuwien.ac.at}
\vspace{0.5cm}
\end{center}
\begin{abstract}
Motivated by the success of the {\nc} scalar Grosse-Wulkenhaar model, a {\nc} $U_\star(1)$ gauge field theory including an oscillator-like term in the action has been put forward in~\cite{Blaschke:2007b}. The aim of the current work is to analyze whether that action can lead to a fully renormalizable gauge model on {\nc} Euclidean space. In a first step, explicit one-loop graph computations are hence presented, and their results as well as necessary modifications of the action are successively discussed.
\end{abstract}
\newpage
\tableofcontents

%%%%%%%%%%%%%%%%%%%%%%%%%%%%%%%%%%%%%%%%%
\section{Introduction}
%%%%%%%%%%%%%%%%%%%%%%%%%%%%%%%%%%%%%%%%%
One of the big challenges of the past decade has been the construction of renormalizable quantum field theories (QFTs) on {\nc} space-time, usually plagued by the infamous {\uim} problem~\cite{Minwalla:1999,Douglas:2001,Rivasseau:2007a}. 
Only in very special cases has this task been successful so far, namely in the case of scalar field theories formulated in even dimensional $\theta$-deformed Euclidean space: In fact, three such renormalizable models are known today\footnote{The other two renormalizable scalar models were introduced in references~\cite{Rivasseau:2008a,Grosse:2008df} and developed further in~\cite{Tanasa:2008a,Blaschke:2008b,Tanasa:2008b,Tanasa:2008c}.}, the first of which being the so-called Grosse-Wulkenhaar model~\cite{Grosse:2003,Grosse:2004b,Rivasseau:2005a}. In that model, the action is supplemented by an oscillator-like term which modifies the propagator in such a way that it becomes Langmann-Szabo invariant~\cite{Szabo:2002} (just like the modified action, in fact). 

Motivated by the success of the Grosse-Wulkenhaar model, some {\nc} gauge field models including similar oscillator-like terms in the action were put forward in Refs.~\cite{Blaschke:2007b,Grosse:2007,Wulkenhaar:2007}. The current work mainly deals with loop calculations based on the previous letter~\cite{Blaschke:2007b}, aiming to clarify whether the action we proposed there can lead to a renormalizable {\nc} gauge theory. In doing so, we find effective one-loop contributions to the action showing great similarities to the so-called ``induced gauge theories'' with oscillator terms which were put forward in~\cite{Grosse:2007,Wulkenhaar:2007}. These findings lead us to the conclusion, that (even though we do not claim to have solved the many problems concerning {\nc} gauge field theories in general, as was discussed in~\cite{Blaschke:2009c}) the implementation of a Grosse-Wulkenhaar oscillator term into gauge theories seems to be a promising ansatz for a renormalizable {\nc} gauge theory. However, it also seems to be necessary to do this in a gauge invariant manner, i.e. to consider an action derived by (or motivated from) the ``induced gauge theories''.

Our paper is organized as follows: Section~\ref{sec:action_and_properties} briefly reviews the action and its symmetries followed by Section~\ref{sec:1loop} where some one-loop results are presented. 
We finally compare the effective action with the induced gauge theory of~\cite{Grosse:2007,Wulkenhaar:2007} in Section~\ref{sec:conclusion} and discuss the consequences.

%%%%%%%%%%%%%%%%%%%%%%%%%%%%%%%%%%%%%%%%%
\section{The {\nc} gauge field model with oscillator term}
\label{sec:action_and_properties}
%%%%%%%%%%%%%%%%%%%%%%%%%%%%%%%%%%%%%%%%%
In the simplest case of $\th$-deformed space-time, one considers a constant commutator $\starco{x^\mu}{x^\nu} \equiv  x^{\mu} \star
x^{\nu} -x^{\nu} \star x^{\mu} = \ri \th^{\mu \nu}$, i.e. a Heisenberg algebra for the space-time coordinates, realized through the so-called Groenewold-Moyal $\star$-product~\cite{Groenewold:1946,Moyal:1949}. Considering four dimensional Euclidean space, in Ref.~\cite{Blaschke:2007b} we proposed the following {\nc} $U_\star(1)$ gauge field action\footnote{Some other candidates for {\nc} gauge field theories without oscillator terms were in fact presented in Refs.~\cite{Slavnov:2003,Blaschke:2008a,Blaschke:2009a,Vilar:2009,Blaschke:2009b}. However, these will not be discussed here.} at
tree-level:
\begin{align}\label{new-action}
\Gam&=\Act_{\text{inv}}+\Act_{\text{m}}+\Act_{\text{gf}}\,,\nonumber\\
\Act_{\text{inv}}&=\inv{4}\intx\, F_{\mu\nu}\star F_{\mu\nu}\,,\nonumber\\
\Act_{\text{m}}&=\frac{\Omega^2}{4}\intx\left(\inv{2}\staraco{\tx_\mu}{A_\nu}\star\staraco{\tx_\mu}{A_\nu}+\staraco{\tx_\mu}{\bc}\star\staraco{\tx_\mu}{c} \right)=\nonumber\\
&=\frac{\Omega^2}{8}\intx\left(\tx_\mu\star\mathcal{C}_\mu\right)\,,\nonumber\\
\Act_{\text{gf}}&=\intx\Bigg[B\star\partial_\mu A_\mu-\inv{2}B\star B-\bc\star\partial_\mu sA_\mu-\frac{\Omega^2}{8}\,\tc_\mu\star s\,\mathcal{C}_\mu\Bigg]\,,
\end{align}
with
\begin{align}
F_{\mu\nu}&=\partial_\mu A_\nu-\partial_\nu A_\mu-\ig\starco{A_\mu}{A_\nu}\,,\nonumber\\
\mathcal{C}_\mu&=\Big(\staraco{\staraco{\tx_\mu}{A_\nu}}{A_\nu}+\starco{\staraco{\tx_\mu}{\bc}}{c}+\starco{\bc}{\staraco{\tx_\mu}{c}}\Big)\,,\nonumber\\
\tx_\mu&=\left(\th^{-1}\right)_{\mu\nu}x_\nu\,.
\end{align}
The field $A_\mu$ denotes the {\nc} generalization of a $U(1)$ gauge field (hence the notation $U_\star(1)$ for the deformed algebra), and $B$ is the multiplier field implementing a non-linear gauge fixing\footnote{Notice, that in the limit $\W\to0$ this becomes a Feynman gauge.}
\begin{align}
\var{\Gam}{B}&=\partial_\mu A_\mu-B+\frac{\Omega^2}{8}\Big(\starco{\staraco{\tx_\mu}{c}}{\tc_\mu}-\staraco{\tx_\mu}{\starco{\tc_\mu}{c}}\Big)=0.
\end{align}
Furthermore, we have an additional multiplier field $\tc_\mu$ imposing on-shell BRST invariance of the expression $\mathcal{C}_\mu$, $\W$ is a constant parameter and $c$, $\bc$ are the ghost/antighost, respectively. The action (\ref{new-action}) is invariant under the BRST transformations given by
\begin{align}\label{BRST}
&sA_\mu=D_\mu c=\partial_\mu c-\ig\starco{A_\mu}{c}, && s\bc=B,\nonumber\\
&sc=\ig{c}\star{c}, && sB=0,\nonumber\\
&s\tc_\mu=\tx_\mu, && s^2\varphi=0\ \forall\ \varphi\in\left\{A_\mu,B,c,\bc,\tc_\mu\right\}.
\end{align}
Using these transformations, the action can be written succinctly as
\begin{align}
\Gamma^{(0)} = \int d^4x \left(
\frac 14 F_{\mu\nu}\star F_{\mu\nu} + s\left( \frac{\Omega^2}8 \tilde c_\mu \star \mathcal{C}_\mu + 
\bar c\star \partial_\mu A_\mu - \frac 12 \bar c \star B \right) \right)\,.
\end{align}
The properties of the appearing fields are summarized in Table 1, where $g_\sharp$ denotes the ghost number.

\begin{table}[!ht]
\caption{Properties of the fields.}
\label{tab:field_prop}
\centering
\begin{tabular}{l c c c c c}
\hline
\hline
Field       & $A_\mu$ & $c$ & $\bar c$ & $\tilde c_\mu$ & $B$ \\[2pt]
\hline
$g_\sharp$  &    0   &  1  &  -1   &    -1      &  0     \\
Mass dim.   &    1   &  0  &   2   &     1      &  2     \\
Statistics  &    b   &  f  &   f   &     f      &  b     \\
\hline
\hline
\end{tabular}
\end{table}

\begin{figure}[ht]
\centering
\includegraphics[scale=0.8]{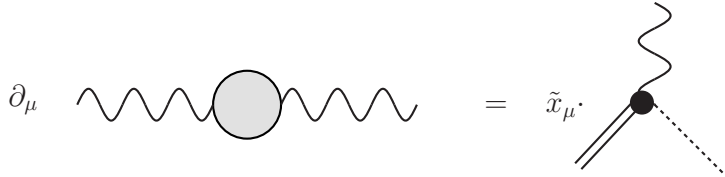}
\caption{Ward identity replacing transversality}
\label{almost_transversality_relation}
\end{figure}

\noindent
By introducing external sources $\rho_\mu$ and $\s$ for the non-linear BRST transformations, one derives the Slavnov-Taylor identity
\begin{align}\label{Slavnov_Taylor_identities}
 \text{S}(\Ga)=\int d^4x
 \,\Bigg(\dif{\Ga}{\rho_\mu}\star\dif{\Ga}{A_\mu}+\dif{\Ga}{\sigma}\star\dif{\Ga}{c}+B\star\dif{\Ga}{\bar{c}}
 +\x_\mu\star\dif{\Ga}{\tilde{c}_\mu}\Bigg)=0\,,
\end{align}
with 
\begin{align}
 \Ga&=\Gamma^{(0)}+\Gamma_{\text{ext}}\,,\nonumber\\
\Gamma_{\text{ext}}&=\intx\left(\rho_\mu \star sA_\mu+ \s \star sc\right)\,.
\end{align}
Equation (\ref{Slavnov_Taylor_identities}) describes the symmetry content with respect to \eqref{BRST}. When taking its functional derivative with respect to $A_\rho$ and $c$ and then setting all fields to zero one arrives at
\begin{align}\label{transversality_relation}
 \pa^z_\mu\vvar{\Ga}{A_\rho(y)}{A_\mu(z)}
 =\int d^4x\,\Bigg(\x_\mu\frac{\delta^3\Ga}{\d c(z)\d A_\rho(y)\d\tilde{c}_\mu(x)}\Bigg)\neq 0\,.
\end{align}
Usually one would expect to obtain the transversality condition for the one-particle irreducible (1PI) two-point graph. However, in this case the oscillator term breaks gauge invariance and hence transversality, as can bee seen from the equation above: Instead one has a Ward identity relating the 1PI two-point graph to a three-point graph involving the new field $\tc_\mu$. Graphically we can express this relation as depicted in Figure~\ref{almost_transversality_relation}.

%%%%%%%%%%%%%%%%%%%%%%%%%%%%%%%%%%%%
The bilinear parts of the action \eqref{new-action} lead to the following gauge field and ghost propagators in momentum space:
\begin{align}\label{newprop}
G^{A}_{\mu\nu}(p,q)&=(2\pi)^4K_M(p,q)\d_{\mu\nu},\nonumber\\
G^{\bc c}(p,q)&=(2\pi)^4K_M(p,q)\,,
\end{align}
where $K_M(p,q)$ denotes the so-called Mehler kernel
\begin{align}\label{Mehler-long-short}
K_M(p,q)&=\frac{\omega^3}{8\pi^2}\int\limits_{0}^{\infty}d\alpha\inv{\sinh^2(\alpha)}\exp\left(-\tfrac{\omega}{4}\coth\left(\tfrac{\alpha}{2}\right)\left(p-q\right)^2-\tfrac{\omega}{4}\tanh\left(\tfrac{\alpha}{2}\right)\left(p+q\right)^2\right),
\end{align}
which is going to be responsible for improving the IR behaviour of the model. In the limit $\omega=\frac{\th}{\Omega}\to\infty$ these propagators 
reduce to the usual ones, i.e.
\begin{align*}
\lim\limits_{\Omega\to0}K_M(p,q)=\inv{p^2}\d^4\left(p-q\right)\,.
\end{align*}
This can be easily verified by integrating over one of the arguments in the Mehler kernel:
\begin{align}
\lim\limits_{\Omega\to0}\int d^4qK_M(p,q)=\lim\limits_{\Omega\to0}\inv{p^2}\left(1-\re^{-\frac{\omega}{2}p^2}\right)=\inv{p^2}\,.
\end{align}
In the following sections we will discuss some one-loop properties of this model, but first some comments are in order:
In contrast to the usual situation, the multiplier field $B$ appears in a $\tc Bc$ vertex (in addition to the $BB$ and $BA$ propagators). However, since the new multiplier $\tc_\mu$ does not propagate, it is impossible to build a graph including these additional Feynman rules, but with only external gauge field legs. In fact, these are the types of graphs we are going to concentrate on in the following, and hence the Feynman rules including the fields $B$ and $\tc$ are ignored for now.

%%%%%%%%%%%%%%%%%%%%%%%%%%%%%%%%%%%%%%%%%
\section{One-loop computations}\label{sec:1loop}
\subsection{Power counting}\label{sec:powercounting}
%%%%%%%%%%%%%%%%%%%%%%%%%%%%%%%%%%%%%%%%%
Before starting the explicit calculations, we may derive estimations for the ``worst case'', i.e. the superficial degree of ultraviolet divergence, which via {\uim} is directly related to the degree of {\nc} IR divergence. 
We take into account the powers of internal momenta $k$ each Feynman rule contributes and also that each loop integral over 4-dimensional space increases the degree by 4. For example, the gauge boson propagator usually behaves like $1/k^2$ for large $k$ and therefore reduces the degree of divergence by 2, whereas each ghost vertex contributes one power of $k$ to the numerator of a graph, hence increasing the degree by one. However, in the present model all propagators, being essentially Mehler kernels, depend on two momenta which additionally need to be integrated in loop calculations. Taking into account these considerations for all  Feynman rules we arrive at
\begin{align}
 d_{\gamma}=4L-6I_A-6I_c-5I_{AB}-4I_B+V_c+V_{3A}+V_{\tc cA}\,,
\end{align}
where the $I$ and $V$ denote the number of the various types of internal lines and vertices, respectively. The number of loop integrals $L$ is given by
\begin{align*}
 L=2I_A+2I_c+2I_{B}+2I_{AB}-(V_c+V_{3A}+V_{4A}+V_{\tc cA}+V_{\tc c2A}+V_{\tc Bc}+V_{\tc\bc2c} -1)\,.
\end{align*}
Furthermore, we take into account the relations
\begin{align}\label{eq:power-counting-relations}
E_{c/\bar c}+2I_c&=2V_c+V_{\tc cA}+V_{\tc c2A}+V_{\tc Bc}+3V_{\tc\bc2c}\,,\nonumber\\ 
 E_A+2I_A+I_{AB}&=V_c+3V_{3A}+4V_{4A}+V_{\tc cA}+2V_{\tc c2A}\,,\nonumber\\
 E_B+2I_{B}+I_{AB}&=V_{\tc Bc}\,,\nonumber\\
 E_{\tilde c}&=V_{\tc cA}+V_{\tc c2A}+V_{\tc Bc}+V_{\tc\bc2c}\,,
\end{align}
between the various Feynman rules describing how they (and how many) can be connected to one another. The $E_{c/\bc}$, $E_A$, $E_{\tc}$ and $E_B$ denote the number of external lines of the respective fields. Using these relations one can eliminate all internal lines and vertices from the power counting formula and arrive at
\begin{align}\label{power-counting_final}
d_\gamma=4-E_A-E_{c/\bar c}-E_{\tilde c}-2E_B\,.
\end{align}
For the gauge boson self-energy we therefore expect the degree of divergence (UV and {\nc} IR) to be at worst quadratically. Gauge invariance usually reduces the degree of UV divergence to be merely logarithmic. However, due to the Ward identity \eqref{transversality_relation} whose right hand side is non-zero, the UV divergence will in fact be worse in our case, namely quadratic.

%%%%%%%%%%%%%%%%%%%%%%%%%%%%%%%%%%%%%%%%%%%%%%%
\subsection{Tadpole graphs}\label{sec:tadpoles}
\begin{figure}[ht]
\centering
\includegraphics[scale=0.65]{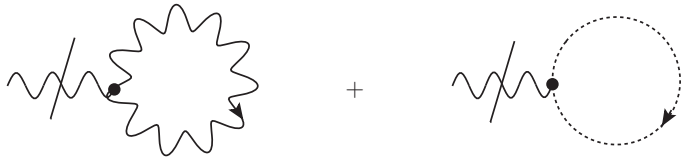}
\caption{tadpole graphs}
\label{fig:tadpoles}
\end{figure}

%%%%%%%%%%%%%%%%%%%%%%%%%%%%%%%%%%%%%%%%%%%%%%%%%
The two possible one-point functions (tadpoles) of this model at one-loop level are depicted in Figure~\ref{fig:tadpoles}.
According to the Feynman rules given in \eqref{Mehler-long-short} and in Appendix~\ref{app:vertices}, the sum of tadpole graphs is given by
\begin{align}
\Pi_\mu(p)&=2\ig\intk \intk'\d^4\left(p+k'-k\right)\sin\left(\frac{k\p}{2}\right)K_M(k,k')\left[2k_\mu+3k'_\mu\right]\,,
\end{align}
where the abbreviation $\p_\mu\equiv\th_{\mu\nu}p_\nu$ has been introduced\footnote{Concerning notation, notice that while in $x$-space we use $\tx_\mu\equiv(\th^{-1})_{\mu\nu}x_\nu$, in momentum space we have $\p_\mu\equiv\th_{\mu\nu}p_\nu$ (and likewise for all other momenta such as $\k_\mu$ or $\q_\mu$).}. In the following, we furthermore assume that the deformation
matrix $( \th_{\mu\nu} )$ has the simple block-diagonal form
\begin{align}
( \th_{\mu\nu} )
=\th\left(\begin{array}{cccc}
0&1&0&0\\
-1&0&0&0\\
0&0&0&1\\
0&0&-1&0
\end{array}
\right) \, ,  \qquad {\rm with} \ \; \th \in \mathbb{R} \, .
\end{align}
Hence one has the identity $\p^2=\th^2p^2$ which will significantly simplify calculations. 
Making use of
\begin{align}
\sin\left(\frac{k\p}{2}\right)=\sum_{\eta=\pm1}\frac{\eta}{2\ri}\exp\left(\frac{\ri\eta}{2}k\p\right)\,,
\end{align}
and considering ``short'' and ``long'' variables defined by $u=k-k'$ and $v=k+k'$ one arrives at
\begin{align}
\Pi^\vare_\mu(p)&=\frac{g\omega^3}{2^{8}\pi^2}\sum\limits_{\eta=\pm1}\int d^4v\left[5v_\mu-p_\mu\right]\int\limits_{\vare}^{\infty}d\a\frac{\eta \re^{\frac{\ri\eta}{4}v\p}}{\sinh^2\a}\exp\left(-\tfrac{\omega}{4}\left[\coth\left(\tfrac{\a}{2}\right)p^2+\tanh\left(\tfrac{\a}{2}\right)v^2\right]\right)\nonumber\\
&=\frac{5\ig\p_\mu}{64}\int\limits_{\vare}^{\infty}d\a\frac{\cosh\left(\frac{\a}{2}\right)}{\sinh^5\left(\frac{\a}{2}\right)}\exp\left[-\inv{4}\coth\left(\frac{\a}{2}\right)\left(\omega+\frac{\th^2}{4\omega}\right)p^2\right]\,,
\end{align}
regularizing the integrals by introducing a UV cutoff $\vare = 1/\Lambda^2$.
Now we consider the following expansion:
\begin{align}\label{expansion-tadpole}
\int \frac{d^4p}{(2\pi)^4}\,\Pi^\vare_\mu(p)&\left[A_\mu(0)+p_\nu\left(\partial^p_\nu A_\mu(p)\big|_{p=0}\right)+\frac{p_\nu p_\rho}{2}\left(\partial^p_\nu\partial^p_\rho A_\mu(p)\big|_{p=0}\right)+\right.\nonumber\\
&\left. \;+\frac{p_\nu p_\rho p_\s}{6}\left(\partial^p_\nu\partial^p_\rho\partial^p_\s A_\mu(p)\big|_{p=0}\right)+\ldots\right]\,.
\end{align}
All terms of even order (i.e. of order 0,2,4,\ldots) are zero for symmetry reasons. Of the other terms, we now show that only the first two, namely orders 1 and 3, diverge in the limit $\vare\to 0$:
\begin{itemize}
 \item \emph{order 1:}
\begin{align}
\int \frac{d^4p}{(2\pi)^4}\,p_\nu\Pi^\vare_\mu(p)&=
\int\limits_{\vare}^{\infty}d\a\frac{5\ig\th_{\mu\nu}}{32\pi^2\omega^3\left(1+\frac{\W^2}{4}\right)^3\sinh^2(\a)}\nonumber\\
&=\frac{5\ig\th_{\mu\nu}}{32\pi^2\omega^3\left(1+\frac{\W^2}{4}\right)^3}\left[\inv{\vare}-1+\mathcal{O}(\vare)\right].
\end{align}
With the external field, we obtain a counter term of the form
\begin{align}
\label{tadpole-order1}
\left(\partial^p_\nu A_\mu(p)\big|_{p=0}\right)& \int \frac{d^4p}{(2\pi)^4}\,p_\nu\Pi^\vare_\mu(p)  =\nonumber\\
&=
\frac{5 g \W^2}{32\pi^2 \omega\left(1+\frac{\W^2}{4}\right)^3} 
\left[ \inv{\vare}-1+\mathcal{O}(\vare) \right] \intx \, \tx_\mu A_\mu(x) \,.
\end{align}
\item \emph{order 3:}
\begin{align}
\int \frac{d^4p}{(2\pi)^4}\, \frac{p_\a p_\b p_\gamma}{6} \Pi^\vare_\mu(p) & = 
\frac {-5\ig \left(\d_{\a\b}\th_{\mu\gamma}+\d_{\b\gamma}\th_{\mu\a}+\d_{\a\gamma}\th_{\mu\b}\right) } {24\pi^2\omega^4 \left(1+\frac{\W^2}{4}\right)^4}
\left[ \ln\vare + \mathcal{O}(0) \right],
\end{align}
and with the external field we get the counter term
\begin{align}
\label{tadpole-order3}
\nonumber
\left(\partial^p_\a \partial^p_\b \partial^p_\gamma A_\mu(p)\big|_{p=0}\right) 
&\int \frac{d^4p}{(2\pi)^4}\, \frac{p_\a p_\b p_\gamma}{6} \Pi^\vare_\mu(p)  =\\
&
= \frac{ 5g } { 8\pi^2 } \frac { \W^4 } { \left(1+\frac{\W^2}{4}\right)^4}
\left[ \ln\vare + \mathcal{O}(0) \right] \intx\, \tx_\mu {\tx}^2 A_\mu (x) 
\,.
\end{align}

\item \emph{order 5 and higher:}\\
These orders are \emph{finite}. The contribution to order $5 + 2n$, $n\ge 0$ is proportional to 
\begin{align}
 \int\limits_{0}^{\infty} d\a \frac{ \sinh^n \frac \a 2 } { \cosh^{n+4} \frac \a 2 } = \frac 4 { (n+1) (n+3) }\,.
\end{align}
\end{itemize}
Notice, that all tadpole contributions would vanish in the limit $\W\to0$ as expected. When keeping $\W\neq0$, the two divergent terms can be removed by renormalization, i.e. by considering the appropriate counter terms given in Eqns.~(\ref{tadpole-order1}) and (\ref{tadpole-order3}), respectively. Remarkably, these terms are present in the induced action calculated in Refs.~\cite{Grosse:2007,Wulkenhaar:2007}. The fact that these graphs do not vanish also means that we need to find the correct vacuum for $\W\neq0$ by solving the equations of motion, which at the classical level read
{\allowdisplaybreaks
\begin{subequations}\label{eom-osci}
\begin{align}
\var{\Gam}{A_\nu}&=\left(-\Delta_4+\Omega^2\tx^2\right)A_\nu+\ig\starco{A_\mu}{F_{\mu\nu}}+\ig\partial_\mu\starco{A_\mu}{A_\nu}+\ig\staraco{\partial_\nu\bc}{c}+\nonumber\\*
&\quad\;\,+\partial_\nu(\partial A)-\partial_\nu B+\frac{\Omega^2}{8}\Big(\staraco{\starco{D_\nu c}{\tc_\mu}}{\tx_\mu}+\starco{\staraco{D_\nu c}{\tx_\mu}}{\tc_\mu}\Big)-\nonumber\\*
&\quad\;\,-\ig\frac{\Omega^2}{8}\Big(\staraco{c}{\staraco{\tx_\mu}{\staraco{A_\nu}{\tc_\mu}}}+\staraco{c}{\staraco{\tc_\mu}{\staraco{\tx_\mu}{A_\nu}}}\Big)=0,\\
\var{\Gam}{B}&=\partial_\mu A_\mu-B+\frac{\Omega^2}{8}\Big(\starco{\staraco{\tx_\mu}{c}}{\tc_\mu}-\staraco{\tx_\mu}{\starco{\tc_\mu}{c}}\Big)=0,\\
\var{\Gam}{\bc}&=\left(-\Delta_4+\Omega^2\tx^2\right)c-\ig\frac{\Omega^2}{8}\Big(\staraco{\staraco{\tx_\mu}{c\star c}}{\tc_\mu}+\staraco{\tx_\mu}{\staraco{\tc_\mu}{c\star c}}\Big)\nonumber\\*
&\quad+\ig\partial_\mu\starco{A_\mu}{c}=0,\\
\var{\Gam}{c}&=\left(\Delta_4-\Omega^2\tx^2\right)\bc+\frac{\Omega^2}{8}\Big(\staraco{\tc_\mu}{\staraco{\tx_\mu}{B}}+\staraco{\tx_\mu}{\staraco{\tc_\mu}{B}}\Big)-\nonumber\\*
&\quad\;\,-\ig\starco{A_\mu}{\partial_\mu\bc}-\frac{\Omega^2}{8}D_\nu\Big(\staraco{\tx_\mu}{\staraco{A_\nu}{\tc_\mu}}+\staraco{\staraco{\tx_\mu}{A_\nu}}{\tc_\mu}\Big)+\nonumber\\*
&\quad\;\,+\ig\frac{\Omega^2}{8}\Big(\starco{c}{\starco{\tc_\mu}{\staraco{\tx_\mu}{\bc}}}-\starco{c}{\staraco{\tx_\mu}{\starco{\bc}{\tc_\mu}}}\Big)=0,\\
\var{\Gam}{\tc_\mu}&=-\frac{\Omega^2}{8}s\,\mathcal{C}_\mu=0.
\end{align}
\end{subequations}
Finding solutions to these equations is the task of a work in progress\footnote{In fact, some work in this respect has been done in Refs.~\cite{Wallet:2007b,Wallet:2008a} for the induced gauge theory case without BRST ghosts.}.
}

%%%%%%%%%%%%%%%%%%%%%%%%%%%%%%%%%%%%%%%%%
\subsection{Two point functions at one-loop level}
\label{sec:one-loop-vac-pol}
\begin{figure}[ht]
\centering
\includegraphics[scale=0.65]{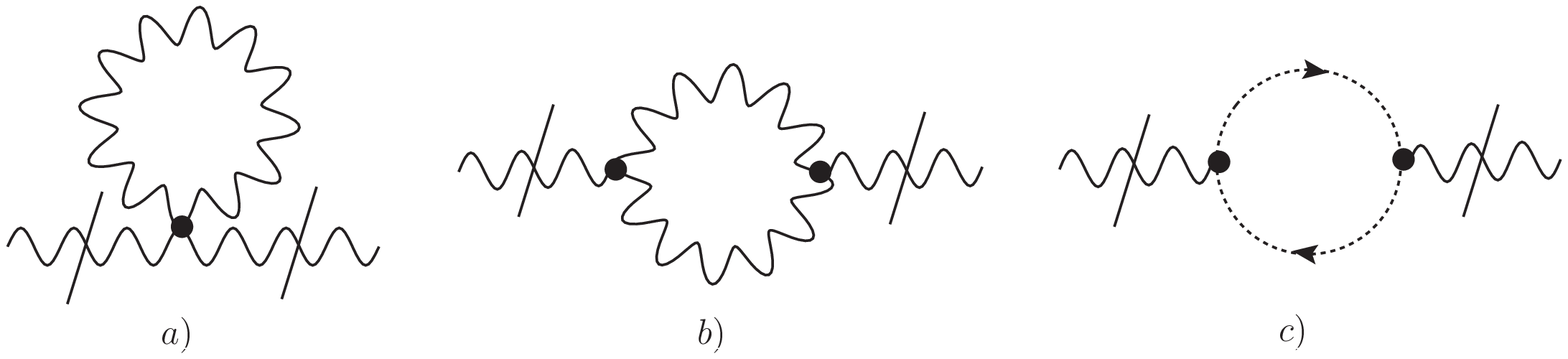}
\caption{Gauge boson self-energy --- amputated graphs}
\label{fig:photonself}
\end{figure}

%%%%%%%%%%%%%%%%%%%%%%%%%%%%%%%%%%%%%%%%%
In this section we analyze the divergence structure of the gauge boson self-energy at one-loop level. The relevant graphs are depicted in Figure~\ref{fig:photonself}.
Explicitly, the sum of these graphs is computed using the same techniques as in the previous section (except for the expansion which is not necessary here). Once more we use long and short variables such as $u=k-k'$ and $v=k+k'$ and including their symmetry factors arrive at the following expressions for the three separate graphs:
{\allowdisplaybreaks
\begin{subequations}
\begin{align}
\Pi^a_{\mu\nu}&=-\frac{3g^2\d_{\mu\nu}}{8}\int d^4v \,K_M(p-p',v)\Big[\sin\left(\tfrac{(v+p')\p}{4}\right)\sin\left(\tfrac{(v+p)\p'}{4}\right)\nonumber\\*
&\quad\hspace{5.3cm}+\sin\left(\tfrac{(v-p')\p}{4}\right)\sin\left(\tfrac{(v-p)\p'}{4}\right)\Big]\,,
\\
\Pi^b_{\mu\nu}&=\frac{g^2}{8}\int d^4u\,d^4v\,K_M(u,v)K_M(u+p-p',v+p+p')\sin\left(\tfrac{(v+u)\p}{4}\right)\sin\left(\tfrac{(v-u)\p'}{4}\right)\nonumber\\*
&\quad\qquad\times\Big[\tfrac{5}{2}(v_\mu v_\nu-u_\mu u_\nu)+\tfrac{3}{2}(u_\mu v_\nu-v_\mu u_\nu)+\tinv{2}p'_\mu(v+u)_\nu+\tinv{2}(v-u)_\mu p_\nu\nonumber\\*
&\quad\qquad\qquad+2p_\mu(v-u)_\mu+2(v+u)_\mu p'_\nu+2p_\mu p'_\nu-4p'_\mu p_\nu \nonumber\\*
&\quad\qquad\qquad+\d_{\mu\nu}\left(\tfrac{v^2-u^2}{2}+\tfrac{p'(v+u)}{2}+\tfrac{(v-u)p}{2}+5pp'\right)\Big]\,,
\\
\Pi^c_{\mu\nu}&=\frac{-g^2}{16}\int d^4u\,d^4v\,K_M(u,v)K_M(u+p-p',v+p+p')\,(v+u)_\mu(v-u+2p')_\nu\nonumber\\*
&\quad\hspace{2.7cm}\times\sin\left(\tfrac{(v+u)\p}{4}\right)\sin\left(\tfrac{(v-u)\p'}{4}\right)\,,
\end{align}
\end{subequations}
where $p$ and $p'$ denote the external momenta. 
In the limit $\W\to0$ one would expect a result $\Pi_{\mu\nu}(p,p')=\Pi_{\mu\nu}(p)\d^4(p-p')$ where the transversality property $p_\mu\Pi_{\mu\nu}=0$ holds. However, due to $\W\neq0$ these properties are not fulfilled, i.e. transversality is broken and one cannot split off a delta function. In order to reveal the divergence structure of the general result without the ``smeared out delta function'', we additionally integrate over $p'$. Finally, noticing that the parameter integrals entering from the Mehler kernels \eqref{Mehler-long-short}, are dominated by the region of small $\alpha$, one also needs to approximate for $\alpha\ll1$ in order to extract the UV and IR divergent terms. We hence arrive at
}
\begin{align}\label{eq:one-loop_divergences}
\Pi_{\mu\nu}^{\text{div}}(p)&=
\frac{g^2 \d _{\mu\nu} \left(1-\tfrac{3}{4}\W^2\right)}{4\pi^2 \omega\,\vare \left(1+\tfrac{\W^2}{4}\right)^3}+
\frac{3 g^2 \delta_{\mu\nu} \W^2}{8\pi^2 \p^2 \left(1+\tfrac{\W^2}{4}\right)^2}+
\frac{2 g^2 \p_\mu \p_\nu }{\pi ^2 (\p^2)^2 \left(1+\tfrac{\W^2}{4}\right)^2}\nonumber\\
&\quad+\text{logarithmic UV divergence}\,.
\end{align}
In the limit $\W\to0$ (i.e. $\omega\to\infty$) this expression reduces to the usual transversal term
\begin{align}
\lim\limits_{\W\to0}\Pi_{\mu\nu}^{\text{div}}(p)&=\frac{2g^2}{\pi^2} \frac{\p_\mu\p_\nu}{(\p^2)^2}+\text{logarithmic UV divergence}\,,
\end{align}
which is quadratically IR divergent\footnote{In fact, this term is consistent with previous results~\cite{Blaschke:2005b,Hayakawa:1999b,Ruiz:2000} calculated in the ``na{\"i}ve'' model, i.e. without any additional non-local terms in the action.} in the external momentum $p$ and logarithmically UV divergent. 
Notice that the general result \eqref{eq:one-loop_divergences}, on the other hand, not only breaks transversality due to the first two terms, but also has an ultraviolet divergence parameterized by $\vare$, whose degree of divergence is higher compared to the (commutative) gauge model without oscillator term. Both properties are due to the term $\Act_{\text{m}}$ in the action which breaks gauge invariance (cf. Eqn. (\ref{transversality_relation})).

%%%%%%%%%%%%%%%%%%%%%%%%%%%%%%%%%%%%%%%%%
\subsection{Vertex corrections at one-loop level}
\label{sec:vertex-corr}
%%%%%%%%%%%%%%%%%%%%%%%%%%%%%%%%%%%%%%%%%
The calculation of the vertex corrections generally proceeds along the lines of the previous section. Due to the vast amount of terms, it is however useful to use a computer: In fact we ``taught'' Wolfram Mathematica$^{\text{\textregistered}}$ to perform exactly the same steps that we would have done by hand.
\begin{figure}[!ht]
 \centering
 \includegraphics[scale=0.8]{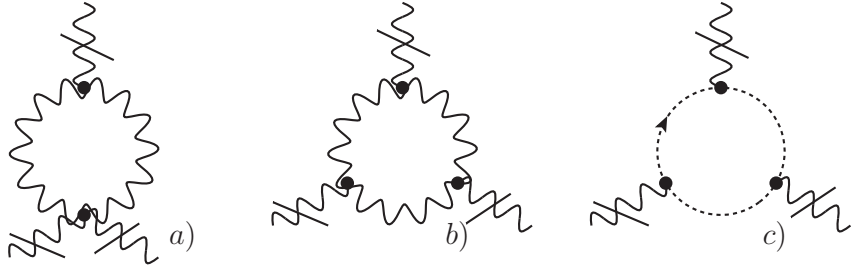}
 \caption{One loop corrections to the 3A-vertex.} 
 \label{fig:1loop_3A_all}
\end{figure}
Hence, computing the graphs depicted in \figref{fig:1loop_3A_all} (and approximating for momentum conservation as in the previous subsections), one eventually finds a linear IR divergence of the form:
\begin{align}\label{eq:3Avert}
 \Gamma^{\text{3A,IR}}_{\mu\nu\rho}(p_1,p_2,p_3)=\frac{-8\ig^3}{\pi^2\left(4+\W^2\right)^3}\sum\limits_{i=1}^{3}\bigg[&
 \frac{16\p_{i,\mu}\p_{i,\nu}\p_{i,\rho}}{\p_i^4}+\frac{3\W^2}{\p_i^2}
 \left(\delta_{\mu\nu}\p_{i,\rho}+\d_{\mu\rho}\p_{i,\nu}+\d_{\nu\rho}\p_{i,\mu}\right)\bigg],
\end{align}
where $p_3=-p_1-p_2$. Once more, this expression is not transversal due to the non-vanishing oscillator term parametrized by $\W$. However, in the limit $\W\to0$ transverality is recovered, and \eqref{eq:3Avert} reduces to the well-known expression~\cite{Matusis:2000jf, Armoni:2000xr,Ruiz:2000}
\begin{align}
 \lim\limits_{\W\to0}V^{\text{1loop}}_{\mu\nu\rho}(p_1,p_2,p_3)=\frac{-2\ig^3}{\pi^2}\sum\limits_{i=1}^{3}\bigg[&
 \frac{\p_{i,\mu}\p_{i,\nu}\p_{i,\rho}}{\p_i^4}\bigg].
\end{align}
In the ultraviolet, the graphs of \figref{fig:1loop_3A_all} diverge only logarithmically.
 
Additionally, one has of course also one-loop corrections to the 4A-vertex $\Gamma^{\text{4A,IR}}_{\mu\nu\rho\s}$. However, these show only a logarithmic divergence, as expected from the power counting (\ref{power-counting_final}).

%%%%%%%%%%%%%%%%%%%%%%%%%%%%%%%%%%%%%%%%%
\section{Discussion}
\label{sec:conclusion}
%%%%%%%%%%%%%%%%%%%%%%%%%%%%%%%%%%%%%%%%%

As already mentioned, the occurring UV counter terms of Section~\ref{sec:1loop} are present in the induced gauge action, e.g.~\cite{Grosse:2007,Wulkenhaar:2007}.
Let us compare the expressions in more detail here. The induced (Euclidean) gauge action (in the notation of~\cite{Grosse:2007}) is given by
\begin{align}
\label{result2}
\nonumber
\Gamma_{I}= \intx\, \bigg\{&
\frac{3}{\th_I} \left(1-\rho_I^2\right) \left(\tilde \mu_I^2-\rho_I^2\right)\left(\tilde X_\nu \star \tilde X_\nu -\x^2\right) 
\\
& 
+ \frac{3}{2}\left(1-\rho_I^2\right)^2 \left( \left(\tilde X_\mu\star \tilde X_\mu\right)^{\star 2}-\left(\x^2\right)^2 \right)
+ \frac{\rho_I^4}{4}  F_{\mu\nu}\star F_{\mu\nu}
\bigg\}\,,
\end{align}
where 
\begin{align}
\rho_I = \frac{1-\W_I^2}{1+\W_I^2},\qquad
\tilde \mu_I^2 = \frac{\mu_I^2\th_I}{1+\W_I^2}\,.
\end{align}
The parameter $\mu_I$ denotes the mass of the  scalar field\footnote{The index $I$ in all variables of \eqref{result2} merely indicate that they belong to the ``induced'' action and need not be equal to the according ones in our present model.}. 
Through its coupling to $A_\mu$, the scalar field ``induced'' the effective one-loop action \eqref{result2} above. The so-called covariant coordinates $\tilde X_\mu$ are furthermore defined as
\begin{align}
 \tilde X_\mu&=\x_\mu+A_\mu\,.
\end{align}
The first expression in the induced action (\ref{result2}) can be written as 
\begin{align}\label{eq:induced-first-term}
\frac{3}{\th_I} \left(1-\rho_I^2\right) \left(\tilde \mu_I^2-\rho_I^2\right)\left(\tilde X_\nu \star \tilde X_\nu -\x^2\right) 
= \frac{3}{\th_I} \left(1-\rho_I^2\right) \left(\tilde \mu_I^2-\rho_I^2\right)\left(2 \x_\nu A_\nu + A_\nu \star A_\nu\right)\,.
\end{align}
The first term of \eqref{eq:induced-first-term} has to be compared with the first expression in (\ref{tadpole-order1}), whereas the second one corresponds to 
the first term of the self energy (\ref{eq:one-loop_divergences}). However, the exact coefficients do not match. But since (\ref{tadpole-order1}) and 
(\ref{eq:one-loop_divergences}) only take one-loop effects into account this cannot be expected. 
Due to technical difficulties, we did not calculate the logarithmic UV divergences in all cases. Therefore, we 
can only compare the term proportional to $\x^2 (\x A)$ given in Eq.(\ref{tadpole-order3}).
The respective term in the induced action --- stemming from the $(\tilde X_\mu \star \tilde X_\mu)^2$ term --- reads 
\begin{align}
6\left(1-\rho_I^2\right)^2 \x^2 \left(\x A\right)\,.
\end{align}

In conclusion, one can state that the induced gauge theory action of Refs.~\cite{Grosse:2007,Wulkenhaar:2007} seems to be the more fundamental one when considering {\nc} gauge theories with Grosse-Wulkenhaar oscillator terms. 
The UV counter terms we have encountered here can be nicely accommodated. 
However, these models exhibit non-trivial vacuum configurations (which are discussed in e.g.~\cite{Grosse:2007jy, Wallet:2008a, Wohlgenannt:2008mn}) due to non-vanishing tadpoles, and it is hence not (yet) clear, how to do higher order loop calculations. Especially, a (highly desirable) general proof of renormalizability will be very involved.

\subsection*{Acknowledgements}
%%%%%%%%%%%%%%%%%%%%%%%%%%%%%%%%%%%%%%%%%%
The authors are indebted to R.~Sedmik for providing valuable assistance with Wolfram Mathematica$^{\text{\textregistered}}$.\\
The work of D.~N.~Blaschke, E. Kronberger and M. Wohlgenannt was supported by the ``Fonds zur F\"orderung der Wissenschaftlichen Forschung'' (FWF) under contracts P20507-N16 and P21610-N16.

%%%%%%%%%%%%%%%%%%%%%%%%%%%%%%%%%%%%%%%%%
\appendix
\section{Vertices}\label{app:vertices}
%%%%%%%%%%%%%%%%%%%%%%%%%%%%%%%%%%%%%%%%%
The vertices used for the graphs in Section~\ref{sec:1loop} are given by
{\allowdisplaybreaks
\begin{subequations}
\begin{align}\label{V3photon}
\wtild{V}^{3A}_{\rho\s\tau}(k_1, k_2, k_3)&=2\ig(2\pi)^4\d^4(k_1+k_2+k_3)\left[(k_3-k_2)_\rho \d_{\s\tau}\right.\nonumber\\*
&\quad\quad\left.+(k_1-k_3)_\s \d_{\rho\tau}+(k_2-k_1)_\tau \d_{\rho\s}\right]\sin\left(\frac{k_1\k_2}{2}\right),
\\\label{V4photon}
\wtild{V}^{4A}_{\rho\s\tau\e}(k_1, k_2, k_3, k_4)&=-4g^2(2\pi)^4\d^4(k_1+k_2+k_3+k_4)\nonumber\\*
&\quad\times\left[(\d_{\rho\tau}\d_{\s\e}-\d_{\rho\e}\d_{\s\tau})\sin\left(\frac{k_1\k_2}{2}\right)\sin\left(\frac{k_3\k_4}{2}\right)\right.\nonumber\\*
&\quad\quad\left.+(\d_{\rho\s}\d_{\tau\e}-\d_{\rho\e}\d_{\s\tau})\sin\left(\frac{k_1\k_3}{2}\right)\sin\left(\frac{k_2\k_4}{2}\right)\right.\nonumber\\*
&\quad\quad\left.+(\d_{\rho\s}\d_{\tau\e}-\d_{\rho\tau}\d_{\s\e})\sin\left(\frac{k_2\k_3}{2}\right)\sin\left(\frac{k_1\k_4}{2}\right)\right],
\\\label{Vghost}
\wtild{V}^{c}_\mu(q_1, k_2, q_3)&=-2\ig(2\pi)^4\d^4(q_1+k_2+q_3)q_{3\mu}\sin\left(\frac{q_1\q_3}{2}\right).
\end{align}
\end{subequations}
}

%%%%%%%%%%%%%%%%%%%%%%%%%%%%%%%%%%%%%%%%%%%%%%%%%%%%%%%%%%

%%%%%%%%%%%%%%%%%%%%%%%%%%%%%%%%%%%%%%%%%%%%%%%%%%%%%%%%%%
% \bibliographystyle{../../custom1.bst}
% \bibliography{./../../articles,./../../books}
\end{document}